\documentclass[fleqn,useAMS,usenatbib]{mnras}
\usepackage{graphicx}	
\usepackage{lineno}
\usepackage{soul}	
\usepackage[T1]{fontenc}
\usepackage{graphicx}	
\usepackage{amsmath}	
\usepackage{amssymb}	

\title[Radial Differential Rotation of Solar Corona]{Radial Differential Rotation of Solar Corona using Radio Emissions}

\author[Vivek Kumar Singh, Satish Chandra, Sanish Thomas, Som Kumar Sharma \& Hari Om Vats]{Vivek Kumar Singh,$^{1}$\thanks{vivek.singh@shiats.edu.in (VKS),} Satish Chandra,$^{2}$\thanks{satish0402@gmail.com (SC),} Sanish Thomas$^{1}$\thanks{sanish.thomas@shiats.edu.in (ST),} Som Kumar Sharma,$^{3}$\thanks{somkumar@prl.res.in (SKS),} and Hari Om Vats$^{4}$\thanks{serfahmedabad@gmail.com (HOV)}\\
$^{1}$Department of Physics, Sam Higginbottom University of Agriculture Technology and  Sciences, Praygraj, India\\
$^{2}$Department of Physics, Pt. Prithi Nath PG College, Kanpur, India\\
$^{3}$Physical Research Laboratory, Ahmedabad, India\\
$^{4}$Space Education and Research Foundation, Ahmedabad, India}

\begin{document}

\date{Accepted XXX. Received YYY; in original form ZZZ}

\pagerange{\pageref{firstpage}--\pageref{lastpage}} 

\pubyear{2020}

\maketitle


\label{firstpage}

\begin{abstract}
The present work is an effort to investigate possible radial variations in the solar coronal rotation by analyzing the solar radio emission data at 15 different frequencies (275-1755 MHz) for the period starting from July 1994 to May 1999. We used a time series of disk-integrated radio flux recorded daily at these frequencies through radio telescopes situated at Astronomical Observatory of the Jagellonian University in Cracow. The different frequency radiation originates from different heights in the solar corona. Existing models, indicate its origin at the height range from nearly $\sim12,000$ km (for emission at 275 MHz), below up to $\sim2,400$ km (for emission at 1755 MHz). There are some data gaps in the time series used for the study, so we used statistical analysis using the Lomb-Scargle Periodogram (LSP). This method has successfully estimated the periodicity present in time series even with such data gaps. The rotation period estimated through LSP shows variation in rotation period, which is compared with the earlier reported estimate using auto correlation technique. The present study indicates some similarity as well as contradiction with studies reported earlier. The radial and temporal variation in solar rotation period are presented and discussed for the whole period analyzed.
\end{abstract}


\section{Introduction}

Since the Sun is a typical middle-age star in the sense of its temperature, brightness and size and is closet (at about 1 astronomical unit) from Earth, which makes Sun as an amazing laboratory for astrophysics. The closeness from Earth marks a very intense signal reaching to Earth in almost all electromagnetic frequency, which allows us to study the solar surface and solar atmospheric features remotely. 

Solar atmosphere pour out radio emission with gradually changing intensity. The radio emissions emerging out of different layers of solar atmosphere have higher frequency in the solar chromosphere in comparison to its corona, and changes slowly with time. This variation is caused due to the change in the sunspot number (SSN) and its groups on the solar surface. The radio waves emanate from solar atmosphere may consist of emissions from three kinds of sources. This may be from the surface of quite Sun, or from the active regions developing on the Sun or the short-lived features developed in the solar atmosphere.

\citet{Aschwanden1995} and \citet{Melendez1999} proposed a theoretical model which gives an estimation of the density, changing with the heights, in the layers above the surface of the Sun. The suggested model gives the plasma frequency of solar atmospheric layers at different heights. According to model suggested by \citet{Benz1993} the dynamic spectra (e.g., from coronal type III solar radio bursts) have two same kind of fine structures, which are displaced by about a factor of two in frequency. This has been explained as the fundamental plasma frequency. According to the altitude-temperature profiles given by \citet{Fontenla1999}, there is a decline in temperature from approx. 6,000 K at the solar surface to the 4,800 K at $\sim500$ km height above surface and then again increases to 6,000 K at $\sim900$ km, about 8,000 K at $\sim1,900$ km, and 10,000 K at  $\sim2,100$ km. Afterward, the temperature starts increasing rapidly to approx. 50,0000 K in very thin transition region beyond 2,100 km, rises up to 90,0000 K at $\sim2,800$ km, and then, reaches up to a million degree temperature or beyond in the solar corona. 

The exact height levels from where these radio frequencies breakout are still unsettled and evaluations of height levels are approximate only. Nevertheless, it is assumed that, emissions at higher frequency would originate from lower heights in comparison to the emissions at lower frequency. Such relative comparison considered to be qualitatively valid, particularly, if the emission data under consideration are averaged over long duration of time, i.e., for years. So, it is valid to assume that radio frequencies originate above the photosphere \citet{Kane2004, Kane2009} at 275 MHz ($\sim12,000$ km), 410 MHz ($\sim10,000$ km), 606 MHz ($\sim4,000$ km), 1,415 MHz ($\sim2,500$ km), escapes from decreasing heights, ranges from outer corona to inner corona. The dependence of spectral features on different parts of solar atmosphere (photosphere to corona) could be studied mainly by studying time series of individual spectral lines known to be emerging in the different regions. Each region of the solar atmosphere is also a basis of radio wave emissions of definite frequencies and their time-series may be helpful for spectral studies. Though the altitudes as mentioned above are fairly accurate, generally, lower frequencies can be assumed to originate from outer regions (higher solar altitudes), while higher frequencies from lower regions (lower solar altitudes)\citep{Vats2001, Kane2004}. In this context, outcome of our study is presented in this paper. 

\begin{figure}
		\includegraphics[width=\columnwidth]{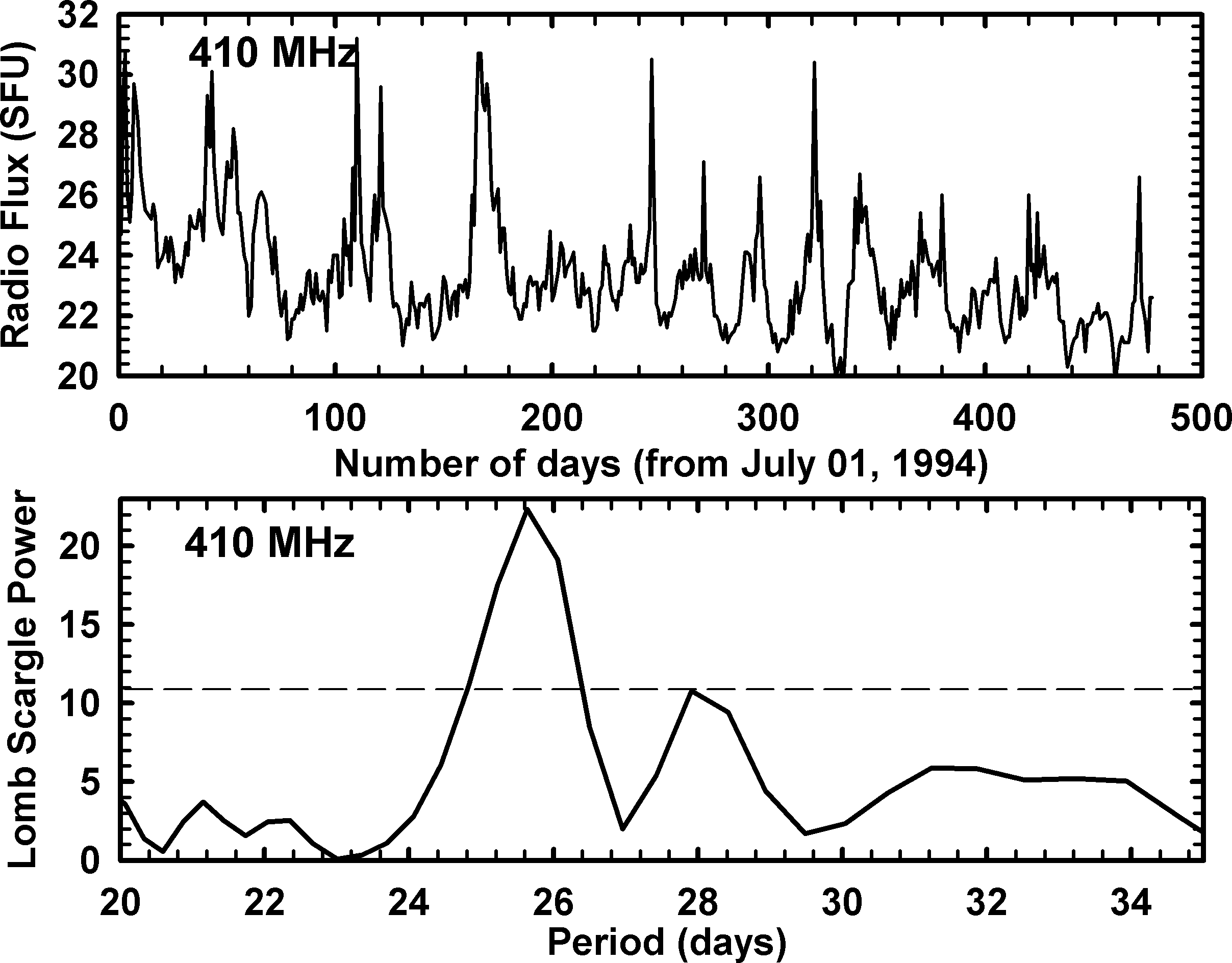}
    \caption{Top Panel: Radio flux at 410 MHz frequency is plotted against days from Jul 01, 1994 to Dec 31, 1996. Bottom Panel: The LSP of radio flux is plotted against period (FAP = 0.05).}
    \label{fig:Figure 1}
\end{figure}

\begin{figure}
		\includegraphics[width=\columnwidth]{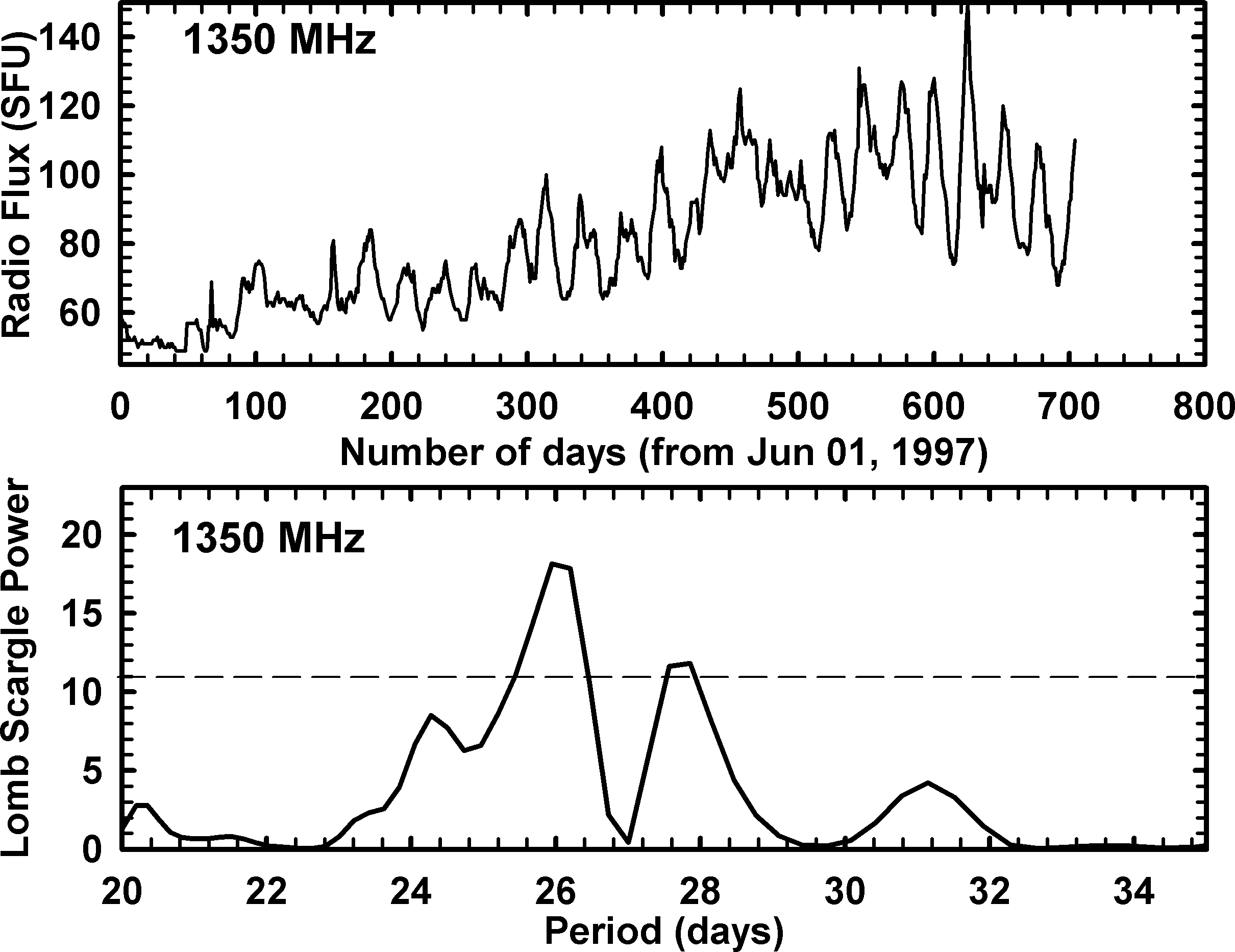}
    \caption{Top Panel: Radio flux at 1350 MHz frequency is plotted against days from Jun 01, 1997 to May 06, 1999. Bottom Panel: The LSP of radio flux is plotted against period (FAP = 0.05).}
    \label{fig:Figure 2}
\end{figure} 

Solar activity indices show significant wide range variation, from few days to few years, on time scales. Long-term changes in solar features (sunspot number) spectral analysis are discussed in detail by \citep{Kane1985, Kane1999, Kane2005, Kane2009}. In recent decades, the analysis was extended to solar UV, EUV, X-ray fluxes and F10 (2800 MHz, 10.7-cm radio) flux \citep{Donnelly1987, Donnelly1990}. \citet{Howard1970} reported that magnetic features and sunspots rotate faster in comparison to the solar surface plasma. This relatively faster rotation of sunspots and magnetic features is attributed to two possible causes; it is either due to its magnetic linkage to  faster rotating layers beneath the solar surface \citep{Howard1984} or perhaps due to the, Coriolis force coupled, buoyancy and drag \citep{D'Silva1994}. \citet{Pap1990} suggested that the coronal X-rays rotated at a period of 28 days, whereas F10 Penticton flux (2.8 GHz) and chromospheric Mg II line rotates in 27 days and also the magnetic and photospheric sources completes one rotation in a period of 25 days. \citet{Vats1998} found prominent periodicity in the variation of the F10 Penticton flux for some years. For example; the reported period was 26.6 days in years 1985-1986 and 1989-1990, whereas 25.8 days in year 1990-1991, but such periodicity lost in other years under investigation. Although, it ranges from 24.1 to 26.4 days, but no systematic relationship found with sunspot numbers. 

For the period 1979 to 1985, synoptic maps of the white-light corona from SOLWIND data were used by \citet{Nash1991} to estimate a synodic rotation rate at 3.5 $R_{\odot}$. Contrary to the photospheric magnetic field, the rotation rate of corona at 3.5 $R_{\odot}$ is found to be rigid, and it matches well with the coronal magnetic field. \citet{Rybak2000} estimated the rotation period of the solar emission corona using Fe XIV 530.3 nm coronal green line measured during 1947 to 1991. For a few years before and after cycle minima, the green corona rotation displays a nearly rigid latitudinal behaviour, but it quickly changes to a differential rotation just after the start of a new solar cycle and lasts until the cycle's descending phase begins. After which a gradual return to the rigid profile occurs. \citet{Sudar2015, Sudar2016} used a segmentation algorithm to detect and track coronal bright points (CBP) in the 19.3 nm channel of an SDO/AIA image and provided a highly accurate solar rotation profile for different periods of time. Whereas, \citet{Wohl2010}, on the other hand, uses 28.4 nm SOHO-EIT images to trace small bright coronal structures (SBCS) to determine the solar differential rotation. They found that the SBCS has a rotation velocity very close to that of small photospheric magnetic features.

\citet{Vats2001} reported that the coronal rotation period at the higher frequency 2800 MHz, which escaped from the lower corona around 60,000 km, is $\sim24.1$ days, which drop off with the ascending height to $\sim23.7$ days at lower frequency (405 MHz), that getaway at $\sim1,30,000$ km height. Whereas, \citet{Bhatt2017} showed the coronal rotation period enhances with drop off in the frequency. The deviation in the period ranges in the duration 24.4 to 22.5 days and the divergence is about three folds more ($\sim1.9$ days) than reported by \citet{Vats2001}.

The work presented here is actually a follow up of the previous studies of rotation period of solar corona \citep{Vats2001, Bhatt2017}, because the finding of both the paper contradicts to each other. The autocorrelation method was used to determine the coronal rotation period in both of the previous studies. We, therefore, are using Lomb Scargle Periodogram (LSP) analysis with the radio emission at various available frequencies to ascertain the coronal rotational period, if any periodicity is present in the signal, and to see any possible systematic variation with the height in solar corona. \citet{Vats2001} and \citet{Bhatt2017} both have used solar radio emission datasets at 275, 405, 670, 810, 925, 1080, 1215, 1350, 1620, 1755, and 2800 MHz (total 11 frequencies) for a period of 26 months (Jun 01, 1997 - Jul 31, 1999), while we analyzed the radio flux at 275, 405, 410, 650, 670, 810, 925, 945, 980, 1080, 1215, 1350, 1450, 1620, \& 1755 MHz (15 frequencies) for the period of about 53 months (Jul 01, 1994 - Dec 31, 1996 and Jun 01, 1997 - May 06, 1999) with a continuous data gap of five months (Figure~3).

\section{Observations and Methodology}

The continuous observations of disc-integrated radio emission measured at 275 - 1755 MHz frequency is presented in this paper are recorded daily through radio telescopes situated at Astronomical Observatory of the Jagellonian University in Cracow, Poland. The radio flux data for period Jul 01, 1994 to 31 Dec 31, 1996 is available at NOAA National Centers for Environmental Information (NCEI). The continuous data from Jun 01, 1997 to May 06, 1999 was available at Cracow Astronomical Observatory (CAO), Poland. The two radio data sets analyzed have data gap of five months (as shown in, Figure~3). The radio emissions at these frequencies evolve mainly due to plasma and free-free bremsstr\"{a}hlung mechanism; which originate from various heights in the solar corona \citep{Raulin2005}.

The variation in the time series of radio flux measurements shows rotational modulation present in the time series (top panels of Figure 1 \& 2). It is, therefore, obvious to use solar flux emitted from a layer of the Sun at certain frequency to estimate the rotation period of that layer of solar atmosphere. This method of estimation of periodic modulation in flux is known as flux modulation method, which can also be employed with the daily measurements of disk-integrated solar emission at different frequency (such as Penticton flux at 2.8 GHz frequency, \citep{Vats1998, Vats2001, Chandra2009, Chandra2011, Li2012, Xie2017}) to measure differential rotation in the solar atmosphere. Although, the time series of flux inherently shows a rotational modulation, even though, determining rotation period directly from variation in flux is not convincingly possible and hence, some time series statistical analysis technique is needed to ascertain synodic rotation period. Generally, the flux/intensity variation in such time series has either ascending or descending inclination due to the raising or falling trend in sunspot cycle, magnetic field reversal cycle and other similar short and long term cycle. It is, therefore, required to use a statistical tool, which can estimate and segregate all the periodic components hidden in time series. 

\begin{figure}
		\includegraphics[width=\columnwidth]{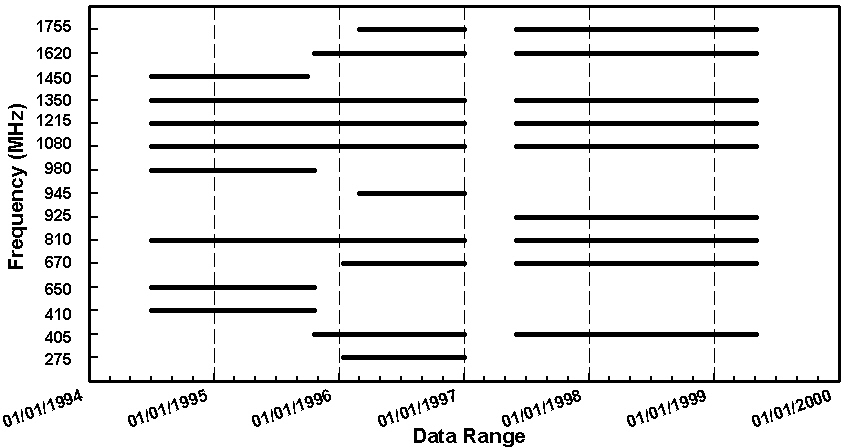}
    \caption{The lines show the duration of data used in the analysis for each frequency in the two data sets (1) 14 frequencies during Jul 01, 1994 to Dec 31, 1996 and (2) 9 frequencies during Jun 01, 1997 to May 06, 1999.}
    \label{fig:Figure 3}
\end{figure}

The LSP is one such time series statistical analysis technique and is a well-known algorithm for detecting and characterizing periodicity with unevenly sampled time series \citep{Lomb1976, Scargle1982, Horne1986, VanderPlas2018}, and has been in wide use within the astronomy and astrophysics community. The LSP is one of such statistical method that offers effective computation of power spectrum estimation, just like in Fourier Transformation, from any data which is sampled unequally. The analysis is a sensitive means of estimating the period of oscillation. The LSP deduces an estimate of the Fourier power with respect to the period of oscillation. The locations of prominent power peaks on the period of oscillation provide the rotation periods of all periodic components present in the time-series. After computing the power of periodogram, the false alarm probability (FAP) criterion is generally needed to see whether it confirms the detection of any periodic signal, or not. Estimating a FAP level of power numerically requires at least the computation of $10/power$ \citep{Delisle2020}. 

The reliability of LSP has been examined by \citet{Li2019} by introducing random data gap of 20\% in a time-series and then period are estimated. The same process, with 2000 different random data gaps, was repeated and periods are determined after each process. The error found in estimation of period was less than 0.2\% established the feasibility of periodogram for estimation of periodicity in a time-series with permissible data gap. The data analysis and results are presented in the upcoming section.

\section{Data Analysis \& Results}

The daily radio flux used in the analysis is the integrated emission over the entire solar disk at a particular frequency which originates at a certain height in solar corona. Continuously measured integrated radio flux for each day constitutes a time-series of radio emission. A strong rotational modulation can be found in such time-series due to the coronal rotation and therefore, it is most appropriate to use statistical method such as LSP to estimate the coronal rotation period from such time-series, even if some random data gap is present in the time series \citep{VanderPlas2018}. For each frequency, two time-series of the radio flux can be generated for the period Jul 01, 1994 to Dec 31, 1996 (876 days) and Jun 01, 1997 to May 06, 1999 (705 days), respectively, but no observations were made for few frequencies in one or another data set. Therefore, out of fifteen radio flux data (275 - 1755 MHz), 9 have two time-series, whereas 6 have only one time-series. When such time-series variation is plotted against days of the period under investigation, the intensity variation shows the dominating periodically varying component throughout the period of study in most of the cases. Typical examples of radio flux intensity variation for both the periods have been plotted in the Figure~1 \& ~2 (top panel). But, at some frequencies the varying component is quite weak and random during some part under investigation. This may be either due to presence of the short lived and weak solar features (like radio noise, radio burst, etc.), or due to the presence of turbulence in the intervening medium, which affected the emissions at these frequencies significantly. 

The LSP analysis is used to find rotation period present in the variation in intensity of radio emission of each time-series. A typical example of such power spectra obtained through LSP is plotted against days in Figure~1 \& ~2 (bottom panel). The position of most prominent peak on days scale provides the estimation of rotation period in power spectrum \citep{Li2019}. The rotation period so obtained is termed as the synodic rotation period, which is a bit higher than actual rotation period because of simultaneous rotation of Earth around the Sun beside Sun's own rotation.  It is, therefore, necessary to introduce correction in the synodic rotation period and convert it into sidereal rotation period, which is actual coronal rotation period of a particular coronal layer \citep{Chandra2011, Li2019}. In this way, we estimated the sidereal coronal rotation period of each frequency ranging from 275 MHz to 1755 MHz for both the data set available (with few exceptions). We also determined the error in the estimation of rotation period for each time series. Sidereal rotation period, so obtained, along with statistical error determined is plotted against the coronal radio frequency for each time series (for 14 radio emissions) observed during July 01, 1994 to Dec 31, 1996 in Figure~4 and same has been plotted in Figure~5 for the period Jun 01, 1997 to May 06, 1999 for 9 radio emissions, all recorded at Astronomical Observatory of the Jagellonian University in Cracow.

\section{Discussion}

The LSP analysis is carried out with time-series of radio flux for 15 closely spaced frequencies between 275 to 1755 MHz, over a period of 53 months in two time span from Jul 01, 1994 to Dec 31, 1996 \& from Jun 01, 1997 to May 06, 1999, to find any systematic variations in coronal rotation with respect to the estimated averaged height above the solar surface.

\begin{figure}
		\includegraphics[width=\columnwidth]{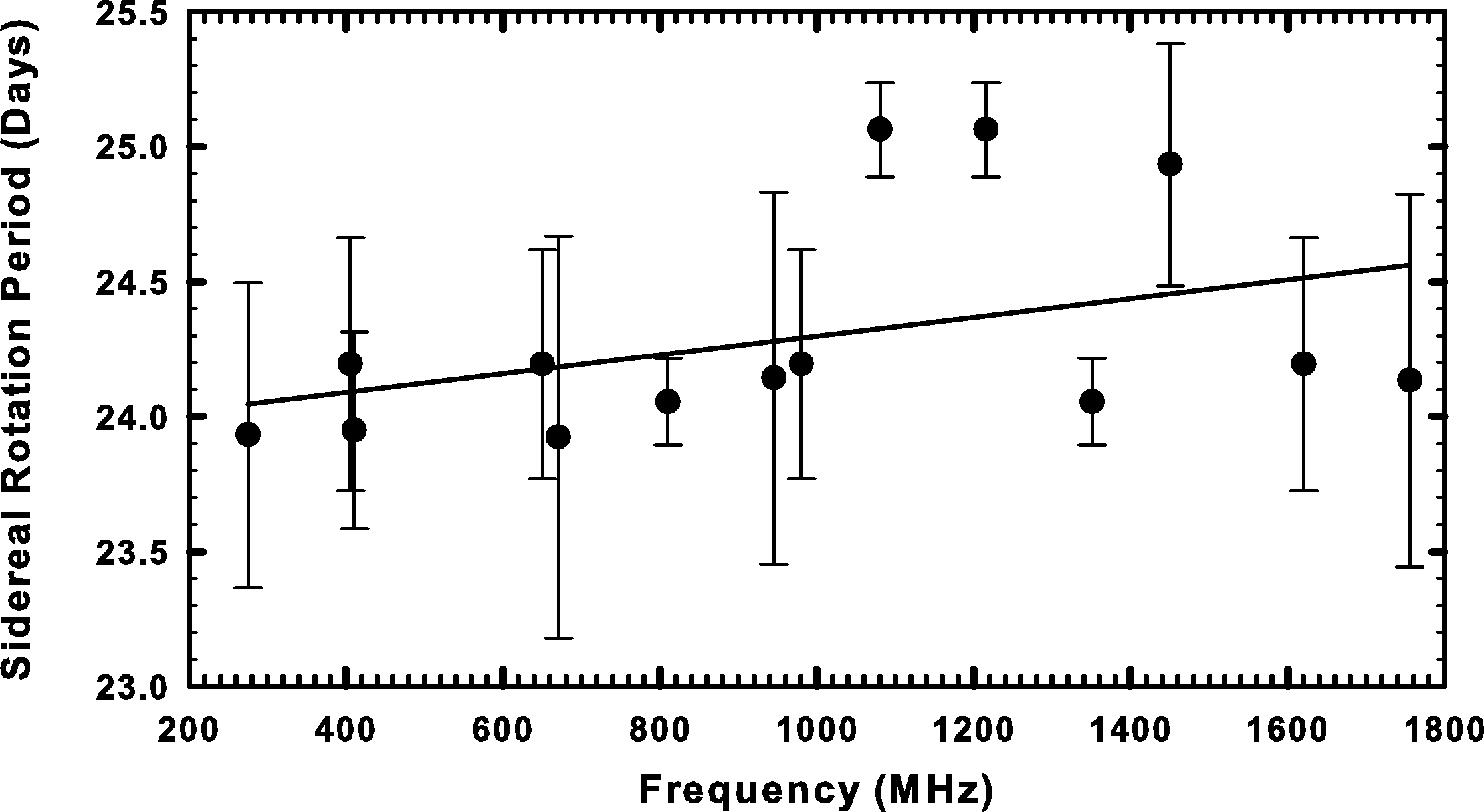}
    \caption{Variation of rotation period as a function of emission frequency (275-1755 MHz) during Jul 01, 1994 to Dec 31, 1996. The slope coefficient is $(+3.47\pm2.33) \times 10^{-4}$ days/MHz}
    \label{fig:Figure 4}
\end{figure}

\begin{figure}
		\includegraphics[width=\columnwidth]{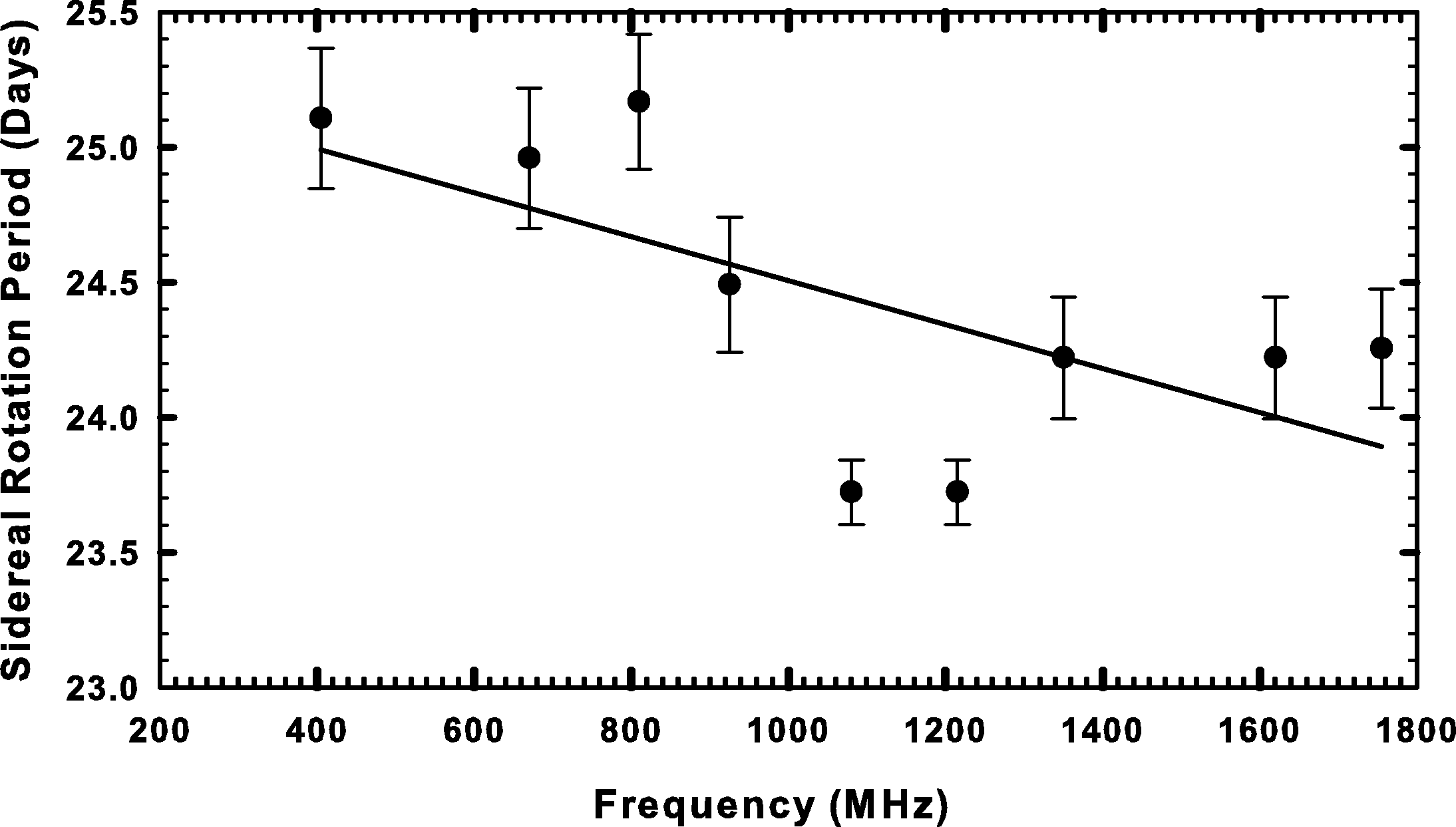}
    \caption{Variation of rotation period as a function of emission frequency (275-1755 MHz) during Jun 01, 1997 to May 06, 1999. The slope coefficient is $(-8.13\pm3.55) \times 10^{-4}$ days/MHz}
    \label{fig:Figure 5}
\end{figure}

The Figure~4 shows an unsystematic variation of sidereal rotation period for coronal emission at 14 different frequencies observed between the span of 30 months from Jul 01, 1994 to Dec 31, 1996 . The variation in period ranges between ~23.93 days (minimum) for 275 MHz to ~25.06 days (maximum) for 1080 \& 1215 MHz. The difference in two extreme rotation period is significant ($\sim 1.25$ days), yet unsystematic in variation. The statistical error is about one day for some of the frequencies due to the availability of lesser number of data point in duration of study; e.g., emissions at 945 \& 1755 MHz (292 days data) or at 275 \& 670 MHz (340 days data), and so on. But, the flux at 810, 1080, 1215, \& 1350 MHz have 876 days data, therefore, uncertainty in the estimation of rotation period at these frequencies are less than 0.3 days. Although the errors in rotation period estimation are overlapping (Figure~4), however, the rotation period for the emissions at closed radio frequencies, such as 405 \& 410, 650 \& 670 and 945 \& 980 MHz, are different from each other. The difference may be due to the different time domain of these observations of radio emissions (as shown in Figure~3). At a single frequency (2.8 GHz) \citet{Vats2010} using autocorrelation method, showed that the coronal rotation period has significant temporal variability.  Thus differences in the rotation period estimates at close by frequencies could be due to temporal variation.

Figure~5 also shows the similar kind of unsystematic variation for emissions at 9 different frequencies recorded in almost 23 months from Jun 01, 1997 to May 06, 1999. The radio emission at 275 MHz has not been included in the Figure~5, because Lomb-Scargle power was well below the false alarm probability (FAP), perhaps due to weak rotational modulation present in the variation of flux at this frequency. The range of variation is from ~23.72 days (minimum) at 1080 MHz to ~25.16 days (maximum) at 810 MHz and difference between maxima and minima is about $\sim1.45$ days. The uncertainty (less than 0.5 day) is almost same for all the emission frequencies, because data was recorded daily for 705 days for all nine frequencies. The \citet{Vats2001} and \citet{Bhatt2017} used the same data for same span of time and estimated the coronal rotation period and shown an increasing and decreasing trend, respectively, in rotation period with respect to coronal emission frequency (Figure~4 in \citet{Bhatt2017}). The sidereal rotation period reported \citep{Vats2001, Bhatt2017} for all 10 frequencies observed at Cracow Astronomical Observatory are being compared with our present investigation (shown in Figure~5) and all the three plots are reconstructed in Figure~6.

The rotation periods deduced in these two studies were mainly based on an autocorrelation analysis with disk-integrated radio flux at 11 different frequencies. As we can see in the Figure~6, \citet{Vats2001} (shown by triangle), who used position of 12th secondary maxima of autocorrelogram on day axis to estimate rotation period, found that rotation period increases with increase in frequency, although the increase is not uniform and systematic for all the frequencies. The variation in rotation periods deviated from 23.6 to 24.15 days, and the divergence was just $\approx0.55$ day. The rotation period estimation accuracy is less than 0.1 days. The trend line fitted in the estimated rotation periods shows increasing trend with slope coefficient of $1.44 \times 10^{-4}$ days/MHz and error in estimation is $\pm 3.76 \times 10^{-5}$ days/MHz. Whereas, \citet{Bhatt2017} (square symbol) used first secondary maxima of autocorrelogram and then Gaussian curve is fitted at its peak as proxy to ascertain the synodic rotation period. Therefore, the uncertainty in estimation of period reduces to one day. The plot in Figure~6 clearly shows the decrease of rotation period with increase in coronal emission frequency. The rotation period found to vary in the range between 24.4 to 22.5 days, with $\approx1.9$ day's difference. In this case, the trend is decreasing with slope of $8.58 \times 10^{-4}$ days/MHz and error $\pm 5.54 \times 10^{-5}$ days/MHz. Hence, both the results are in qualitative disagreement with each other.

\begin{figure}
		\includegraphics[width=\columnwidth]{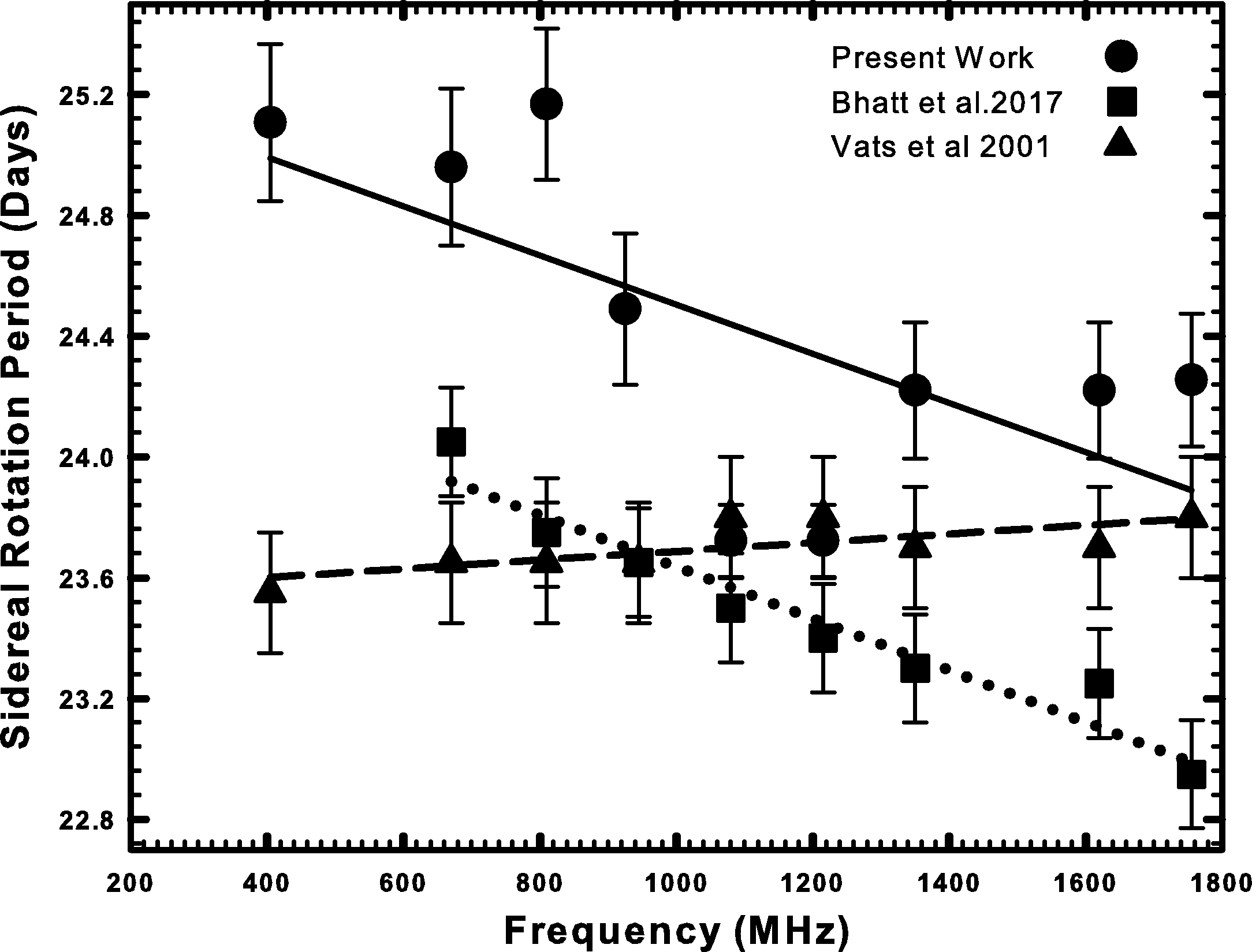}
    \caption{Comparison of our study on coronal rotation with the work of \citet{Bhatt2017} \& \citet{Vats2001} on radio emissions (275-1755 MHz) recorded during Jun 01, 1997 to May 06, 1999.}
    \label{fig:Figure 6}
\end{figure}

When we compared both the reported studies with our present work (symbol of circle in Figure~6) using LSP, we found that although the rotation period calculated through LSP analysis is decreasing with the increase in radio frequency but the decrement is not as uniform or systematic as it has been shown in the plots of \citet{Vats2001, Bhatt2017}. The slope coefficient of trend line is $8.13 \times 10^{-4}$ days/MHz and associated error is $\pm 3.55 \times 10^{-4}$ days/MHz. Although the present study is not in good quantitative agreement with other two reported work, even though, decline in rotation period with frequency of emission is qualitatively in agreement with \citet{Bhatt2017} and non-uniformity in variation of period resembles with \citet{Vats2001}. 

It is interesting to report that rotation periods estimated at 1080 and 1215 MHz (as shown in Figure~6) by present method and by \citet{Vats2001} are very close to each other. The rotation periods reported by \citet{Bhatt2017} are slightly lower. The estimates at higher frequencies and lower frequencies by present work are lower than those reported by both. The general trend of variation of rotation period with frequency is like that reported by \citet{Bhatt2017}.

\section{Conclusions}

This work is based on two data sets of multi-frequency radio observations at Astronomical Observatory Cracow using LSP method to estimate sidereal rotation period: (1) July 1, 1994 to Dec 31, 1996, however, this data has several gaps. Data availability at different frequencies also vary (as seen in Figure~3). This analysis shows that average sidereal rotation period increases with increasing frequency. The average sidereal rotation period changes by about half day. (2) The other data set is for the period Jun 1, 1977 to May 6, 1999. This data is regular and is available for 9 frequencies (405, 670, 810, 925, 1030, 1215, 1350, 1620 and 1755 MHz). This shows that the average sidereal rotation period decreases with increasing frequency.  The average sidereal rotation period changes by about one day. These are in good agreement with results of \citet{Bhatt2017} using autocorrelation method. However, the sidereal rotation periods estimated by LSP method (present work) are higher (overestimated $\sim 1$ day) in comparison with the estimates of \citet{Bhatt2017} using autocorrelation method. From this work, it can be concluded that  the estimates of solar rotation period by LSP method are reasonable. This method works even when the time series has data gaps. Furthermore, the frequency variation of the sidereal rotation period seems to have temporal variability.

\section*{Acknowledgments}

The research is supported by CSJM University, Kanpur, India under its MRP financial grant scheme. We would also like to thanks the anonymous referees for very careful review of our paper and for the comments, corrections and suggestions that ensued.

\section*{Data Availability}

The data underlying this article are available in {\it NGDC} at \url{ https://ngdc.noaa.gov/stp/space-weather/solar-data/solar-features/solar-radio/noontime-flux/cracow} and in {\it  CAO} at \url{http://www.oa.uj.edu.pl/sol/index.html}.


\bsp

\label{lastpage}


\begin{thebibliography}{99}

\bibitem[\protect\citeauthoryear{Aschwanden \& Benz}{1995}]{Aschwanden1995} 
Aschwanden M.J., \& Benz A.O., 1995, \apj, 438, 997

\bibitem[\protect\citeauthoryear{Benz}{1993}]{Benz1993}
Benz A.O., Plasma Astrophysics: Kinetic Proceses in Solar \& Stellar Corona, Kluwer Academic Publishers, Dordrecht, 1993, Vol. 184, 127

\bibitem[\protect\citeauthoryear{Bhatt et al.}{2017}]{Bhatt2017}
Bhatt H., Trivedi R., Sharma .K., \& Vats H.O., 2017, \solphys, 292(4), 55

\bibitem[\protect\citeauthoryear{Chandra, Vats \& Iyer}{Chandra et al.}{2009}]{Chandra2009}
Chandra S., Vats H.O., \& Iyer K.N., 2009, \mnras, 400(1), L34

\bibitem[\protect\citeauthoryear{Chandra, \& Vats}{2011}]{Chandra2011}
Chandra S., \& Vats H.O., 2011, \mnras, 414(4), 3158

\bibitem[\protect\citeauthoryear{Delisle, Hara \& Segransan}{2020}]{Delisle2020}
Delisle, J.-B., Hara, N., \& Segransan, D., 2020, \aap, 635, A83

\bibitem[\protect\citeauthoryear{Donnelly}{1987}]{Donnelly1987}
Donnelly, R. F., 1987, \solphys, 109, 37

\bibitem[\protect\citeauthoryear{Donnelly \& Puga}{1990}]{Donnelly1990}
Donnelly, R.F., \& Puga, L.C., 1990, \solphys, 130, 369

\bibitem[\protect\citeauthoryear{D'Silva \& Howard}{1994}]{D'Silva1994}
D'Silva, S., \& Howard, R.F., 1994, \solphys, 151, 213

\bibitem[\protect\citeauthoryear{Fontenla et al.}{1999}]{Fontenla1999}
Fontenla, J., White, O.R., Fox, P.A., Avrett, E.H., \& Kurucz, R.L., 1999, \apj, 518, 480

\bibitem[\protect\citeauthoryear{Horne \& Baliunas}{1986}]{Horne1986}
Horne, J.H., \& Baliunas, S.L., 1986, \apjs, 302, 757

\bibitem[\protect\citeauthoryear{Howard}{1984}]{Howard1984}
Howard, R., 1984, \araa, 22, 131

\bibitem[\protect\citeauthoryear{Howard \& Harvey}{1970}]{Howard1970}
Howard, R.F., \& Harvey, J., 1970, \solphys, 12, 23

\bibitem[\protect\citeauthoryear{VanderPlas}{2018}]{VanderPlas2018}
VanderPlas, J.T., 2018, \apjs, 236(1),28

\bibitem[\protect\citeauthoryear{Kane \& Trivedi}{1985}]{Kane1985}
Kane, R.P., \& Trivedi, N. R., 1985, J. Geomag. Geoele., 1071, 37

\bibitem[\protect\citeauthoryear{Kane}{1999}]{Kane1999}
Kane, R.P., 1999, \solphys, 189, 217

\bibitem[\protect\citeauthoryear{Kane}{2004}]{Kane2004}
Kane, R.P., 2004, \solphys, 219, 357

\bibitem[\protect\citeauthoryear{Kane}{2005}]{Kane2005}
Kane, R.P., 2005, \solphys, 277, 155

\bibitem[\protect\citeauthoryear{Kane}{2009}]{Kane2009}
Kane, R.P., 2009, Annales Geophysicae, 27(4), 1469

\bibitem[\protect\citeauthoryear{Li et al.}{2012}]{Li2012}
Li, K.J., Shi, X.J., Feng, W. et al., 2012, \mnras, 423, 3584

\bibitem[\protect\citeauthoryear{Li et al.}{2019}]{Li2019}
Li, K.J., Xu, J.C., Yin, Z.Q., \& Feng, W., 2019, \apj, 875(2), 90

\bibitem[\protect\citeauthoryear{Lomb}{1976}]{Lomb1976}
Lomb, N.R., 1976, \apss, 39, 447

\bibitem[\protect\citeauthoryear{Melendez et al.}{1999}]{Melendez1999}
Melendez, J.L.M., Sawant, H.S., Fernandes, F.C.R., \& Benz, A. O., 1999, \solphys, 187, 77

\bibitem[\protect\citeauthoryear{Nash}{1991}]{Nash1991}
Nash, A.G., 1991, \apj, 366, 592

\bibitem[\protect\citeauthoryear{Pap, Tobiska, \& Bouwer}{1990}]{Pap1990}
Pap, J., Tobiska, W.K., \& Bouwer, S.D., 1990, \solphys, 129, 165

\bibitem[\protect\citeauthoryear{Raulin \& Pacini}{2005}]{Raulin2005}
Raulin, J.P., \& Pacini, A.A., 2005, Adv. in Sp. Res., 35, 739

\bibitem[\protect\citeauthoryear{Ryb\'{a}k}{2000}]{Rybak2000} 
Ryb\'{a}k, A., 2000, Hvar Obs. Bull., 24, 135

\bibitem[\protect\citeauthoryear{Scargle}{1982}]{Scargle1982}
Scargle, J.D., 1982, \apj, 263, 835

\bibitem[\protect\citeauthoryear{Sudar et al.}{2015}]{Sudar2015}
Sudar, D., Skoki\'{c}, I., Braj\v{s}a, R., \& Saar, S.H., 2015, \aap, 575, A63

\bibitem[\protect\citeauthoryear{Sudar et al.}{2016}]{Sudar2016}
Sudar, D., Saar, S.H., Skoki\'{c}, I., Beljan, I.P., \& Braj\v{s}a, R., 2016, \aap, 587, A29

\bibitem[\protect\citeauthoryear{Vats et al.}{1998}]{Vats1998}
Vats, H.O., Deshpande, M.R., Shah, C.R., \& Mehta, M., 1998, \solphys, 181, 351

\bibitem[\protect\citeauthoryear{Vats et al.}{2001}]{Vats2001}
Vats, H.O., Cecatto, J.R., Mehta, M., Sawant, H.S., \& Neri, J.A.C.F., 2001, \apjl, 548, L87

\bibitem[\protect\citeauthoryear{Vats et al.}{2010}]{Vats2010}
Vats, H.O., Chandra, S., \& Iyer, K.N., Astrophysics \& Space Science Proceeding, Springer, Berlin, 2010, 526

\bibitem[\protect\citeauthoryear{W\"{o}hl et al.}{2010}]{Wohl2010}
W\"{o}hl, H., Braj\v{s}a, R., Hanslmeier, A., \& Gissot, S.F., 2010, \aap, 520, A29

\bibitem[\protect\citeauthoryear{Xie, Shi \& Zhang}{2017}]{Xie2017}
Xie, J.L., Shi, X.J., \& Zhang, J., 2017, \apj, 841, 42



\end{thebibliography}
\end{document}